# Masking Transmission Line Outages via False Data Injection Attacks

Xuan Liu, *Member, IEEE,* Zhiyi Li, *Student Member, IEEE*, Zuyi Li, *Senior Member, IEEE*

*Abstract*—Today's power systems become more prone to cyber-attacks due to the high integration of information technologies. In this paper, we demonstrate that the outages of some lines can be masked by injecting false data into a set of measurements. The success of the topology attack can be guaranteed by making that: 1)the injected false data obeys KCL and KVL to avoid being detected by the bad data detection program in the state estimation; 2)the residual is increased such that the line outage cannot be detected by PMU data. A quadratic programming problem is set up to determine the optimal attack vector that can maximize the residual of the outaged line. The IEEE 39-bus system is used to demonstrate the masking scheme.

## Nomenclature

| | |
|---|---|
| $M$ | sufficiently large value for line power flow |
| $r$ | residual value |
| $\tau$ | given maximum percentage of change for load measurement attack |
| $b, d\ i, j,$ | subscript: index for buses |
| $l, k$ | subscript: index for lines |
| $ND$ | number of loads |
| $S$ | set of PMU buses |
| $SD$ | set of load buses |
| $SL$ | set of lines |
| $SD_{pmu}$ | set of buses with PMUs |
| $SL_{pmu}$ | set of lines covered by PMUs |
| $\Delta D_i$ | false data injected into the measurement at bus $i$ |
| $\Delta F_l$ | false data injected into the measurement on line $l$ |
| $f_k^0$ | power flow of line $k$ before attacks |
| $\Delta f_k$ | additional power flow of line $k$ due to false data injection |
| $\Delta p, \Delta p'$ | additional bus power injection vector with/without false data injection |
| $\Delta P_i$ | total false power injection at bus $i$ |
| $\boldsymbol{B}$ | bus susceptance matrix |
| $\boldsymbol{B}_{out}$ | bus susceptance matrix after line outage |
| $\boldsymbol{D}$ | load vector |
| $\Delta \boldsymbol{D}$ | false data injection vector into load measurements |
| $\boldsymbol{F}$ | line flow vector |
| $\Delta \boldsymbol{F}$ | false data vector of line measurements |
| $\boldsymbol{H}$ | Jacobian matrix |
| $\boldsymbol{P}_{pre}$ | power injection vector before attacks |
| $\boldsymbol{P}^{inj}$ | power injection vector |
| $\boldsymbol{P}_{post,k}$ | equivalent power injection vector after line $k$ outaged without false data |
| $\boldsymbol{P}'_{post,k}$ | equivalent power injection vector after line $k$ outaged with false data |
| $\boldsymbol{S}$ | shift factor matrix of the power grid |
| $\boldsymbol{S}_k$ | row $k$ of $\boldsymbol{S}$ |
| $\boldsymbol{U}$ | bus-generator incidence matrix |
| $\boldsymbol{V}$ | bus-load incidence matrix |
| $\boldsymbol{W}$ | bus-line incidence matrix |
| $\boldsymbol{z}$ | measurement vector |
| $\Delta \boldsymbol{z}$ | measurement error injected by the attacking vector |
| $\boldsymbol{\theta}$ | phase angle vector |
| $\Delta \boldsymbol{\theta}$ | incremental phase angle vector |
| $\boldsymbol{\theta}_{pre}$ | phase angle vector before attacks |
| $\boldsymbol{\theta}_{post,k}$ | phase angle vector after line $k$ outaged without false data injection |
| $\boldsymbol{\theta}'_{post,k}$ | phase angle vector after line $k$ outaged with false data injection |
| $\Delta \boldsymbol{\theta}_{m,k}$ | observed phase angle change vector of PMU buses after line $k$ is outaged |
| $\Delta \boldsymbol{\theta}_{m,k}^{cal}$ | calculated phase angle change vector of PMU buses after line $k$ is outaged |
| $\Delta \boldsymbol{\theta}_{m,k}^{E}$ | extended vector of $\Delta \boldsymbol{\theta}_{m,k}$ |
| $\Delta \boldsymbol{\theta}'_{out,k}$ | bus phase angle change vector after line $k$ is outaged with false data injection |

Note that $\Delta$ represents incremental change and symbols in bold represent vectors or matrices.

## I. Introduction

POWER grid, one of the critical infrastructures, is the backbone of a nation's economy and is critical to the homeland security. In particular, the blackouts happened around the globe in recent years have raised a great concern about the reliable and safe operation of power systems. In fact, some blackouts are triggered by the initial failures of a small number of components [1]. However, the control center was not aware of the potential risk or even did not detect such failures, thus no corrective measures were taken to mitigate the risk. As a result, the failures of the components triggered the failures of more components and finally led to the blackout. From the perspective of the system operator, it is essential to monitor the real-time operation state of a power system and detect the failures of components in time. To achieve this goal, an increasing number of traditional sensors as well as PMUs are being installed to collect bus voltages, line flows and etc., and then transmit these data into the control center via communication networks.

On the other side, the tight integration between information communication technologies and the physical components in power systems makes power systems more vulnerable to cyber attacks. Consequently, an attacker has a chance to compromise the data generated by the Supervisory Control and Data Acquisition (SCADA) systems to confuse or misguide the state estimation. Liu *et al.* in [2] demonstrated that these attacks against state estimation can be undetectable by injecting the pre-designed false data into meters if the full topology and parameter information of a power grid is assumed to be known. Based on the corrupted data, the control center may make wrong decisions that lead to economic loss or insecure operations. Driven by the pioneering work in [2], extensive researches have been done to investigate the impacts of false data injection attacks on the economic and reliable operation of power systems [3]-[15]. In particular, we in [14] and [15] showed that an attacker only needs to obtain the network information of the attacking region instead of that of the entire power network. This is done by making sure that the variations of phase angles of all boundary buses connected to the same island of the non-attacking region are the same.

Recently, several works have been done to reveal that an attack can also change the real-time topology of a power grid that is sent to the control center. Following the same principle in [2], the authors in [16] showed that the topology information sent to the control center can be attacked. They proposed a state-preserving model for a single line attack, in which a pair of additional powers is injected into the power measurements at the terminal buses of the attacked line. To overcome the practical issues in [16], we proposed a practical topology attack model [17] in which the attacking amount at a bus is limited within a certain range and a heuristic algorithm was proposed to minimize the efforts of obtaining the network information for constructing a feasible attack vector. Li *et al.* in [18] minimized the required network information by replacing the external network with its equivalent network based on the corresponding measurements. This type of topology attacks could bring in serious consequences to the operation of power systems. In particular, a wise attacker will choose a perfect attacking time, for example, when a system is under stressed condition and the outage of a critical line might lead to cascading failures. If such an outage is masked by an attacker and the control center has no way to detect the outage of the line, no actions will be taken and catastrophic consequences may follow. Thus, the security of the topology information of a power grid is worth to pay much attention.

However, the proposed topology attack model in [16]-[18] is only suitable for the power grid without PMUs deployed for line outage detection. Since a set of PMUs have been deployed in power systems, it is necessary for the defender to reinvestigate the possibility of topology attacks by considering the protection role of PMUs. When a line is outaged, the bus phase angles at the PMU buses will change, which allow a defender to detect the outage of the line by utilizing the advantages of PMU data. In [19], the authors proposed a single line outage detection model based on the known network information and PMU data at a subset of buses. In their model, an enumeration approach was used to compute the responding residual value for each line. If the residual of a line is less than a predetermined threshold value, then this line is thought to be outaged. Considering the current costs of deploying PMUs are still expensive, the authors in [20] further explored the optimal placement strategy for PMUs. The objective is to minimize the number of PMUs to be deployed which maximizing the distance differences between every two-line outage combination.

Thus, to achieve an undetectable attack, two conditions must be met: (1) the injected false data can avoid the bad data detection in state estimation; (2) the line outage cannot be detected by PMU data sent to the control center. In this paper, we will investigate the possibility of making line outages by injecting false data. Our focus is to mask the outage of a single line. The main contributions of this paper are three-fold:

(1) We for the first time demonstrate that line outage can be masked by disrupting the PMU data based outage detection by injecting false data. The principle is to increase the residual between the observed PMU angle changes and the calculated angle changes at the PMU buses.
(2) We derive the analytical expression for the residual due to false data and mathematically explain the principle of increasing the residual by injecting false data. A quadratic programming problem is proposed to determine the optimal attack vector that maximizes the residual of the outaged line.
(3) Our work reveals the vulnerability of the real-time topology to cyber-attacks and thus highlights the necessity for a defender to develop some effective protection strategies and corresponding detection methods.

The rest of this paper is organized as follows. Section II reviews the concept of bad data attacks. Section III proposes a masking scheme to hide the outage of a single line. Section IV demonstrates the proposed model with the IEEE 39-bus system. Section V concludes the paper.

## II. REVIEW OF FALSE DATA INJECTION ATTACKS

In the transmission network, the bus voltages are around the rating voltage and the phase angle differences of the terminal buses of a line is small. These characteristics allow us to describe the power flows using the linear Direct Current (DC) model. Based on DC power flow, we have

$$F = X^{-1}W^T \theta \quad (1)$$
$$P^{inj} = B\theta \quad (2)$$

where matrix $B$ is the susceptance matrix

$$B = WX^{-1}W^T \quad (3)$$

Combining (1)-(3) into a compact form, we get

$$z = H\theta \quad (4)$$

where

$$z = \begin{bmatrix} P^{inj} \\ F \\ -F \end{bmatrix} \quad H = \begin{bmatrix} WX^{-1}W^T \\ X^{-1}W^T \\ -X^{-1}W^T \end{bmatrix}$$

In DC state estimation, the state vector $\hat{\theta}$ is estimated by the least square method

$$\hat{\boldsymbol{\theta}} = min\,||\boldsymbol{z} - \boldsymbol{H\theta}||_2 \quad (5)$$

where $\boldsymbol{z}$ is the vector of measurements, and $\boldsymbol{H}$ is the Jacobian matrix of the power grid.

To check the validity of the transmitted data, the control center will perform bad data test which calculates the 2-norm value, or residue, of the error vector between the real measurements and corresponding estimated values. The residue $r$ is

$$r = \left\|\boldsymbol{z} - \boldsymbol{H}\hat{\boldsymbol{\theta}}\right\|_2 \quad (6)$$

If the residue is less than a given threshold value, the estimated state $\hat{\boldsymbol{\theta}}$ is acceptable. Otherwise, bad data is thought to be existing. Thus, while performing the attacks, the attacker has to avoid being detected by bad data test. In particular, if the injection false data vector $\Delta\boldsymbol{z}$ and the state variation vector $\Delta\boldsymbol{\theta}$ satisfy $\Delta\boldsymbol{z} = \boldsymbol{H}\Delta\boldsymbol{\theta}$, then we have

$$\left\|\boldsymbol{z}_a - \boldsymbol{H}\hat{\boldsymbol{\theta}}_{bad}\right\|_2 = \left\|\boldsymbol{z} + \Delta\boldsymbol{z} - \boldsymbol{H}(\hat{\boldsymbol{\theta}} + \Delta\boldsymbol{\theta})\right\|_2 =$$
$$\left\|\boldsymbol{z} - \boldsymbol{H}\hat{\boldsymbol{\theta}} - (\Delta\boldsymbol{z} - \boldsymbol{H}\Delta\boldsymbol{\theta})\right\|_2 = \left\|\boldsymbol{z} - \boldsymbol{H}\hat{\boldsymbol{\theta}}\right\|_2 \quad (7)$$

That is, the residue $r$ will not increase, so false data injection attacks on measurements can bypass the residual test.

The disadvantage of the general false data injection model is the lack of considerations of practical conditions of real power systems. For example, a strong mutual communication is usually built between a power plant and control center, so there is a high risk of being detected if an attacker chooses to attack the measurement at a generator bus. Additionally, the control center would doubt the data if the injecting false data is too large since the state estimator has the pre-knowledge of the load distribution of a power network and can predict loads using load forecasting techniques.

To make the false data injection attack model more practical, Yuan *et al.* in [3] proposed a load redistribution attack model which sets some constraints on the general attack model: (1) The output reading of a generator cannot be altered; (2) The readings of the measurement at a load bus can be attacked within certain ranges. The mathematical model can be formulates as

$$\sum_{d=1}^{ND} \Delta D_d = 0 \quad (8)$$
$$-\tau D_d \leq \Delta D_d \leq \tau D_d (0 < \tau < 1) \quad (9)$$
$$\Delta \boldsymbol{F} = -\boldsymbol{S} \cdot \boldsymbol{V} \cdot \Delta \boldsymbol{D} \quad (10)$$

Constraint (8) ensures that summation of injection powers at load buses equals to zero since the readings of generators cannot be changed. Constraint (9) limits the attacking amount within a certain range. Constraint (10) constructs the corresponding attacking vector of line measurements.

III. MATHEMATICAL FORMULATION

In this section, we first propose the masking principle of line outages, and then formulate a quadratic programming problem to determine the optimal attack vector.

*A. Principle of masking line outages*

In power systems, the topology of a power grid provides the basis for economic operations and security controls. However, it is not invariable due to natural or malicious attacks or intentional line switching. Thus, to monitor the real-time topology of a power network, it is essential for the control center to detect the outages of lines.

In practice, when a line is physically disconnected, 0 will be sent to the control center to represent the line is disconnected. However, an attacker aims to stealthily modify this information from 1 to 0 so that the control center believes the line is still connected.

To ensure that such topology attack will not be detected by the state estimator, as discussed in [14], the injected false data should obey KCL and KVL. When a line $k$ is disconnected, the topology of the power grid will change. Accordingly, the power flow is determined using a new shift factor $\boldsymbol{S}_{out}$. However, consider the objective of an attacker is to mask the outage of this line. That is, the power flow should be consistent with the case that line $k$ is still connected. So, the power flow is determined using the original shift factor $\boldsymbol{S}$ before the line outage. In this case, the received data will be trusted by the control center, the outage of this line is masked.

As discussed in Section I, the deployment of PMUs enables a defender to detect the outages of some lines based on PMU data. Thus, the success of such a topology attack not only needs the injected false data pass the bad data detection procedure in the state estimation, but also requires that the injected data can ensure the outage cannot be detected by PMU data. Next, we first introduce the general mechanism of line outage detection using PMU phasor data.

Before the outage, the phase angle vector is calculated by

$$\boldsymbol{\theta}_{pre} = \boldsymbol{B}^{-1}\boldsymbol{P}_{pre} \quad (11)$$

Similarly, the post-outage phase angle vector is calculated by

$$\boldsymbol{\theta}_{post,k} = \boldsymbol{B}_{out}^{-1}\boldsymbol{P}_{pre} \quad (12)$$

As shown in Fig. 1, the line outage can be simulated by injecting additional powers $\Delta p, -\Delta p$ at the terminal buses of the outaged line while keeping the topology of the power network unchanged [21]. That is, the outaged line is still assumed to be in service. By doing so, the recalculation of the matrix $\boldsymbol{B}_{out}^{-1}$ for each outaged line can be avoided.

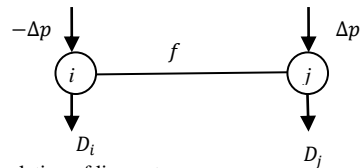

Fig. 1 Simulation of line outage

Accordingly, the post-outage power injection vector becomes

$$\boldsymbol{P}_{post,k} = \boldsymbol{P}_{pre} - \boldsymbol{e}\Delta p \quad (13)$$

where vector $\boldsymbol{e}$ is defined as

$$\boldsymbol{e} = \begin{bmatrix} 0 \\ 0 \\ \vdots \\ 1 \\ \vdots \\ -1 \\ \vdots \\ 0 \end{bmatrix} \begin{matrix} \\ \\ \leftarrow i \\ \\ \leftarrow j \\ \\ \end{matrix}$$

The additional power injection $\Delta p$ is calculated by (14)
$$\Delta p = -\gamma f_k^0 \qquad (14)$$
where $f_k^0$ is the pre-outage flow of line $k$ and $\gamma$ is calculated as follows.
$$\gamma = \frac{x_k}{[(\boldsymbol{B}^{-1})_{ii} + (\boldsymbol{B}^{-1})_{jj} - 2(\boldsymbol{B}^{-1})_{ij}] - x_k}$$
Accordingly, (12) becomes
$$\boldsymbol{\theta}_{post,k} = \boldsymbol{B}^{-1}\boldsymbol{P}_{post,k} \qquad (15)$$
Then, the calculated phase angle changes at PMU buses are given by (16)
$$\Delta\boldsymbol{\theta}_{m,k}^{cal} = (\boldsymbol{\theta}_{post,k})_s - (\boldsymbol{\theta}_{pre})_s \qquad (16)$$
where the subscript $s$ represents only the rows in $\boldsymbol{\theta}_{post,k}$ and $\boldsymbol{\theta}_{pre}$ corresponding to the PMU buses are selected to calculate $\Delta\boldsymbol{\theta}_{m,k}^{cal}$.

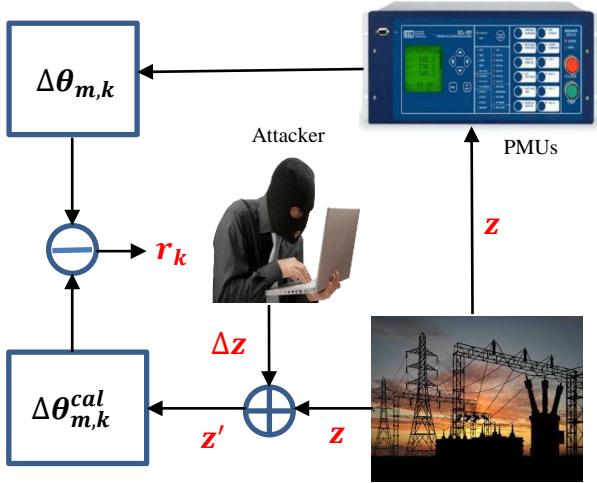

Fig. 2 Principle of masking line outage detection

To detect if there is line outage and which line is on outage, we assume that line $k$ is out of service and then calculate
$$r_k = \min_{f_k^0} \|\Delta\boldsymbol{\theta}_{m,k} - \Delta\boldsymbol{\theta}_{m,k}^{cal}\|_2 \qquad (17)$$
where $\Delta\boldsymbol{\theta}_{m,k}$ is the observed phase angle change vector at PMU buses before and after the outage of line $k$. The original line flow $f_k^0$, which is before the outage of line $k$, is allowed to be variable to achieve the best match [19].

As shown in figure 2, without false data injection, the residual $r_k$ in (17) should be small since $\Delta\boldsymbol{\theta}_{m,k}$ and $\Delta\boldsymbol{\theta}_{m,k}^{cal}$ are determined according to similar system conditions. It has been proved that the outage of a single line can be effectively detected based on the residual between $\Delta\boldsymbol{\theta}_{m,k}$ and $\Delta\boldsymbol{\theta}_{m,k}^{cal}$. The candidate line whose outage results in the minimum residual value is determined as the outaged line. That is, for each candidate line $l$, we calculate its corresponding residual $r_l$ according to (17). The outaged line $k$ is determined by examining the values of the residuals:
$$k = \{l: r_l = \min\{r_l, l = 1,2,\dots,NL\}\} \qquad (18)$$
Line $k$ that has the minimum residual value is identified as the outaged line.

Notice that $\Delta\boldsymbol{\theta}_{m,k}$ comes from the actual readings of PMUs and thus represents the change of PMU phase angles before and after the line outage. In comparison, $\Delta\boldsymbol{\theta}_{m,k}^{cal}$ is calculated using (17) according to the received measurements from remote sensors. This allows an attacker to control $\Delta\boldsymbol{\theta}_{m,k}^{cal}$ by compromising the measurements transmitted from the power grid due to the vulnerability of communication networks. As shown in Fig. 2, if an attacker injects false data $\Delta z$, then $\Delta\boldsymbol{\theta}_{m,k}^{cal}$ is calculated using the corrupted data $z' = z + \Delta z$. But $\Delta\boldsymbol{\theta}_{m,k}$ still comes from the readings of PMUs that follow the system condition without false data injection. In this case, $\Delta\boldsymbol{\theta}_{m,k}$ and $\Delta\boldsymbol{\theta}_{m,k}^{cal}$ are determined based on two different system conditions, making the resulting residual large. As a result, the outage of this line cannot be detected if its residual is increased to be greater than the minimum one. From the perspective of an attacker, in order to mask the outage of line, the goal is to increase the residual such that it is not the minimum one among the residuals of all candidate lines.

### B. Construction of attack vector

In this section, we will present the optimization model for determining the optimal attack vector.

As mentioned before, $\Delta\boldsymbol{\theta}_{m,k}^{cal}$ is calculated based on DC power flow, so it obeys the superposition law. If there is a false data vector $\Delta\boldsymbol{D}$ injected into the power grid, then $\Delta p$ in (13) becomes
$$\Delta p' = -\gamma(f_k^0 + \Delta f_k) \qquad (19)$$
where $\Delta f_k$ is attributed to the false injection data $\Delta\boldsymbol{D}$
$$\Delta f_k = -\boldsymbol{S}_k \cdot \boldsymbol{V} \cdot \Delta\boldsymbol{D} \qquad (20)$$
Under the disruption of injected false data, the equivalent post-outage power injection vector is
$$\boldsymbol{P}'_{post,k} = \boldsymbol{P}_{pre} - \boldsymbol{V} \cdot \Delta\boldsymbol{D} - \boldsymbol{e}\Delta p' \qquad (21)$$
Then, the phase angle vector after line $k$ is outaged can be calculated as
$$\boldsymbol{\theta}'_{post,k} = \boldsymbol{B}^{-1}\boldsymbol{P}'_{post,k} \qquad (22)$$
According to the principle of line outage detection discussed in Section A, to mask the outage of line $k$, an attacker needs to maximize the residual in (17),
$$\max \|\Delta\boldsymbol{\theta}_{m,k} - \Delta\boldsymbol{\theta}_{m,k}^{cal}\|_2 \qquad (23)$$
Define a diagonal matrix $\boldsymbol{R}$ with all entries corresponding to the PMU buses being 1. Extend vector $\Delta\boldsymbol{\theta}_{m,k}$ to an $N \times 1$ column vector $\Delta\boldsymbol{\theta}_{m,k}^E$ with all entries for PMU buses being from $\Delta\boldsymbol{\theta}_{m,k}$ and those for non-PMU set to zeros.

Then, we have
$$\|\Delta\boldsymbol{\theta}_{m,k} - \Delta\boldsymbol{\theta}_{m,k}^{cal}\|_2 = \|\boldsymbol{R}(\Delta\boldsymbol{\theta}_{m,k}^E - \Delta\boldsymbol{\theta}'_{out,k})\|_2 \qquad (24)$$
where $\Delta\boldsymbol{\theta}'_{out,k}$ is calculated by combining (11) and (22),
$$\Delta\boldsymbol{\theta}'_{out,k} = \boldsymbol{\theta}'_{post,k} - \boldsymbol{\theta}_{pre} = \boldsymbol{B}^{-1}\boldsymbol{P}'_{post,k} - \boldsymbol{B}^{-1}\boldsymbol{P}_{pre}$$
$$= \boldsymbol{B}^{-1}(\boldsymbol{P}'_{post,k} - \boldsymbol{P}_{pre}) \qquad (25)$$
Introducing (21) into (25) yields
$$\boldsymbol{\theta}'_{post,k} - \boldsymbol{\theta}_{pre} = \boldsymbol{B}^{-1}(-\boldsymbol{V} \cdot \Delta\boldsymbol{D} - \boldsymbol{e}\Delta p') \qquad (26)$$
Together with (19) and (20), we get

$$\theta'_{post,k} - \theta_{pre} = B^{-1}(-V \cdot \Delta D - e\gamma(f_k^0 + \Delta f_k))$$
$$= B^{-1}[-V \cdot \Delta D - e\gamma(f_k^0 - S_k \cdot V \cdot \Delta D)]$$
$$= B^{-1}[(\gamma e \cdot S_k \cdot V - V)\Delta D] - \gamma f_k^0 B^{-1} \cdot e \quad (27)$$

Then,
$$\Delta \theta^E_{m,k} - \Delta \theta'_{out,k} = \Delta \theta^E_{m,k} + \gamma f_k^0 B^{-1} \cdot e - B^{-1}[(\gamma e \cdot S_k \cdot V - V)\Delta D] \quad (28)$$

Thus,
$$\left\|\Delta \theta_{m,k} - \Delta \theta^{cal}_{m,k}\right\|_2 = \left\|-R \cdot B^{-1}(\gamma e \cdot S_k \cdot V - V)\Delta D + e_1(\Delta \theta^E_{m,k} + \gamma f_k^0 B^{-1} \cdot e)\right\|_2 \quad (29)$$

Define
$$\alpha = -R \cdot B^{-1}(\gamma e \cdot S_k \cdot V - V)$$
$$\beta_1 = R \cdot \Delta \theta^E_{m,k}$$
$$\beta_2 = \gamma R \cdot B^{-1} \cdot e$$

Then, we have
$$r_k = \left\|\Delta \theta_{m,k} - \Delta \theta^{cal}_{m,k}\right\|_2 = \|\alpha \cdot \Delta D + \beta_1 + f_k^0 \beta_2\|_2 \quad (30)$$

To minimize $r_k$ is equivalent to minimize
$$L = (\alpha \cdot \Delta D + \beta_1 + f_k^0 \beta_2)^T(\alpha \cdot \Delta D + \beta_1 + f_k^0 \beta_2)$$
$$= (\alpha \cdot \Delta D + \beta_1)^T(\alpha \cdot \Delta D + \beta_1) + 2f_k^0 \beta_2^T(\alpha \cdot \Delta D + \beta_1) + (f_k^0)^2 \beta_2^T \cdot \beta_2 \quad (31)$$

Calculate the derivative with respect to $f_k^0$, which gives
$$\frac{\partial L}{\partial f_k^0} = 2\beta_2^T(\alpha \cdot \Delta D + \beta_1) + 2f_k^0 \beta_2^T \cdot \beta_2 = 0 \quad (32)$$

Solving the equation, we have
$$f_k^0 = -\frac{\beta_2^T(\alpha \cdot \Delta D + \beta_1)}{\beta_2^T \cdot \beta_2} \quad (33)$$

Introducing (33) into (30), we obtain
$$r_k = \left\|\alpha \cdot \Delta D - \frac{\beta_2^T \cdot \alpha \cdot \Delta D \cdot \beta_2}{\beta_2^T \cdot \beta_2} + \left(\beta_1 - \frac{\beta_2^T \cdot \beta_1 \cdot \beta_2}{\beta_2^T \cdot \beta_2}\right)\right\|_2 \quad (34)$$

Since $\beta_2^T \cdot \alpha$ is a row vector, $\beta_2^T \cdot \alpha \cdot \Delta D$ and $\beta_2^T \cdot \beta_2$ are scalars, the following transformation can be verified.

$$\frac{\beta_2^T \cdot \alpha \cdot \Delta D \cdot \beta_2}{\beta_2^T \cdot \beta_2} = \begin{bmatrix} \beta_{2,1} \frac{\beta_2^T \cdot \alpha \cdot \Delta D}{\beta_2^T \cdot \beta_2} \\ \beta_{2,2} \frac{\beta_2^T \cdot \alpha \cdot \Delta D}{\beta_2^T \cdot \beta_2} \\ \vdots \\ \beta_{2,ND} \frac{\beta_2^T \cdot \alpha \cdot \Delta D}{\beta_2^T \cdot \beta_2} \end{bmatrix} = \begin{bmatrix} \left(\beta_{2,1} \frac{\beta_2^T \cdot \alpha}{\beta_2^T \cdot \beta_2}\right)\Delta D \\ \left(\beta_{2,2} \frac{\beta_2^T \cdot \alpha}{\beta_2^T \cdot \beta_2}\right)\Delta D \\ \vdots \\ \left(\beta_{2,ND} \frac{\beta_2^T \cdot \alpha}{\beta_2^T \cdot \beta_2}\right)\Delta D \end{bmatrix} = G \cdot \Delta D \quad (35)$$

where the matrix
$$G = \begin{bmatrix} \beta_{2,1} \frac{\beta_2^T \cdot \alpha}{\beta_2^T \cdot \beta_2} \\ \beta_{2,2} \frac{\beta_2^T \cdot \alpha}{\beta_2^T \cdot \beta_2} \\ \vdots \\ \beta_{2,ND} \frac{\beta_2^T \cdot \alpha}{\beta_2^T \cdot \beta_2} \end{bmatrix} \quad (36)$$

We further define
$$K = \beta_1 - \frac{\beta_2^T \cdot \beta_1 \cdot \beta_2}{\beta_2^T \cdot \beta_2}$$

Then, we have
$$r_k = \|(\alpha - G) \cdot \Delta D + K\|_2 \quad (37)$$

This formula clearly shows that the residual can be attributed to two parts: the injected false data $\Delta D$ and the term $K$. Without the injection of false data,
$$r_k = \|K\|_2 \quad (38)$$

From the perspective of an attacker, the goal is to construct an optimal attack vector $\Delta D$ to maximize the resulting residual $r_k$. We rewrite (37) as
$$r_k = \sqrt{\frac{1}{2}\Delta D^T \cdot Q \cdot \Delta D + q \cdot \Delta D + K^T \cdot K} \quad (39)$$

where
$$Q = 2(\alpha - G)^T(\alpha - G)$$
$$q = 2K^T(\alpha - G)$$

To ensure the constraints of attacking amounts, for the terminal buses of the target line,
$$-\tau D_i \leq \Delta P_i = -\Delta p' - \Delta D_i \leq \tau D_i \quad (40)$$
$$-\tau D_j \leq \Delta P_j = \Delta p' - \Delta D_j \leq \tau D_j \quad (41)$$
where $\Delta p'$ is the equivalent power injections into the terminal buses used to simulate the line outage, $\Delta P_i$ and $\Delta P_j$ are the total false power injections at bus $i$ and bus $j$, respectively.

Introducing (19) into (40) and (41), which gives
$$-\tau D_i + \gamma f_k^0 \leq \gamma S_k \cdot V \cdot \Delta D - \Delta D_i \leq \tau D_i + \gamma f_k^0 \quad (42)$$
$$-\tau D_j - \gamma f_k^0 \leq -\gamma S_k \cdot V \cdot \Delta D - \Delta D_j \leq \tau D_j - \gamma f_k^0 \quad (43)$$

For the remaining load buses, we have
$$-\tau D_d \leq \Delta D_d \leq \tau D_d \quad \forall d \in SD, d \neq i, d \neq j \quad (44)$$

Considering the false data $\Delta D$ and applying the superposition law, we have
$$\Delta F = S(-V \cdot \Delta D - e(-\gamma \Delta f_k)) \quad (45)$$

Substituting (20) into (45), we have
$$\Delta F = (-S \cdot V - \gamma S \cdot e \cdot S_k \cdot V)\Delta D \quad (46)$$

After this, the optimization problem of determining the optimal attack vector $\Delta D$ to maximize the residual is formulated as a quadratic programming problem (47):
$$\max \frac{1}{2}\Delta D^T \cdot Q \cdot \Delta D + q \cdot \Delta D \quad (47)$$

subject to

Constraints (42)-(44), (46), (9)
$$\Delta D_d = 0 \quad \forall d \in SD_{pmu} \quad (48)$$
$$\Delta F_l = 0 \quad \forall l \in SL_{pmu} \quad (49)$$

The objective function maximizes the residual after false data $\Delta D$ is injected. Constraints (48)-(49) ensures that if a PMU is installed at a bus, then the power injection at the bus and the power flows of lines connected to the bus cannot be attacked. Note that the objective function is quadratic and the constraints are all linear. So, it can be solved by available quadratic programming solvers.

## IV. CASE STUDY

In this section, we test the masking scheme using the IEEE 39-bus system [22], which is composed of 39 buses and 46 lines as shown in Fig. 3. We make the following assumptions

- This system is fully measured. That is, we need one meter to measure the injection power for each bus and two meters to measure the power flow passing through each transmission line.
- Five PMUs are installed at buses 4, 13, 18, 23 and 24, respectively. Once a PMU is installed at a bus, the power injection and flows of branches connected to the buses are measured. Thus, the power injection measurements at buses 4, 18, 23, 24 and line flow measurements on lines 4-5, 4-14, 3-4, 10-13, 13-14, 22-23, 23-24, 23-36, 3-18, 17-18 and 16-24 cannot be attacked.
- An attacker can obtain the network information of the entire power network.

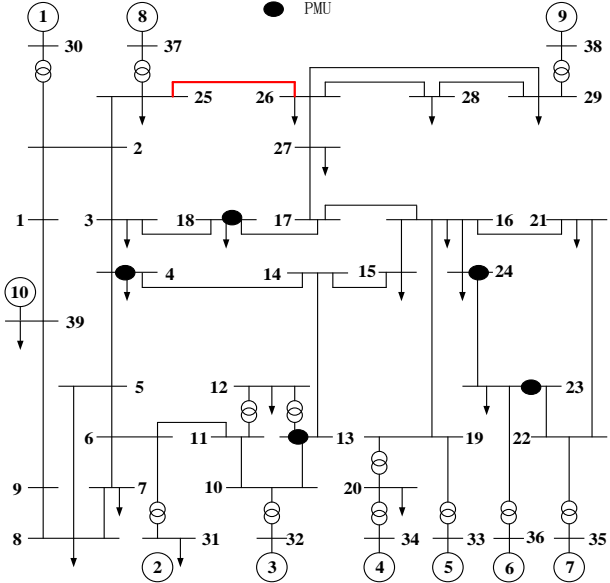

Fig. 3 IEEE 39-bus system

In addition to those lines whose flows are monitored by PMUs, we also exclude from consideration those lines whose outages will lead to the disconnection of the entire power network, as no detection approach is reported to detect the outages of such lines. Hence, the remaining 24 lines are then considered for testing the proposed masking scheme.

First, the outage of line 25-26 (i.e., $k$) is used to demonstrate the principle of the proposed masking scheme in detail. The PMU angle change vector before and after the line outage $\Delta\boldsymbol{\theta}_{m,k}$ is simulated using MATPOWER 4.1 based on the Alternating Current (AC) power flow model [22]. This is because the angle from the PMU measurement represents the actual phase angle at a bus in a real-world system that is described by the AC power flow. The original phase angle before the attack is calculated by (11),

$$\boldsymbol{\theta}_{pmu}^{pre} = \begin{bmatrix} \theta_4 \\ \theta_{13} \\ \theta_{18} \\ \theta_{23} \\ \theta_{24} \end{bmatrix} = \begin{bmatrix} -12.63° \\ -8.93° \\ -11.99° \\ -3.38° \\ -9.91° \end{bmatrix} \quad (50)$$

Then, we disconnect line 25-26 and recalculate the power flow using MATPOWER. The post-outage phase angles at PMU buses become

$$\boldsymbol{\theta}_{pmu}^{post} = \begin{bmatrix} \theta_4 \\ \theta_{13} \\ \theta_{18} \\ \theta_{23} \\ \theta_{24} \end{bmatrix} = \begin{bmatrix} -12.70° \\ -9.04° \\ -12.31° \\ -3.81° \\ -10.36° \end{bmatrix} \quad (51)$$

So, the changes in the PMU phase angles due to the outage of line 25-26 are

$$\Delta\boldsymbol{\theta}_{m,k} = \begin{bmatrix} \Delta\theta_4 \\ \Delta\theta_{13} \\ \Delta\theta_{18} \\ \Delta\theta_{23} \\ \Delta\theta_{24} \end{bmatrix} = \begin{bmatrix} -0.07° \\ -0.11° \\ -0.32° \\ -0.43° \\ -0.44° \end{bmatrix} \quad (52)$$

PMU data are assumed to be noise-free due to its high accuracy. Without loss of generality, the standard deviations of non-PMU measurements are set to 1% of the actual values. Table 1 ranks the top 5 lowest residuals determined by (17) without false data injection. We can see that the outaged line 25-26 has the lowest residual 0.0346, less than half of that of the line ranked the second. Thus, the outage of line 25-26 can be detected using the detection method based on the PMU data.

TABLE 1 TOP 5 LOWEST RESIDUALS BEFORE ATTACK

| rank | Line | $r_k$ |
|---|---|---|
| 1 | 25-26 | 0.0346 |
| 2 | 1-2 | 0.1012 |
| 3 | 16-19 | 0.1022 |
| 4 | 2-3 | 0.1330 |
| 5 | 5-8 | 0.1569 |

Then, we solve the optimization problems (47) to get the false injection data into the measurements at load buses. The maximum attacking amount at a load bus is no more than 50% of the load (i.e., $\tau=0.5$). For line 25-26, we have

$$\gamma = \frac{0.0323}{(0.0381 + 0.0516 - 2 \times 0.0339) - 0.0323} = -3.1582$$

The actual flow of line 25-26 is 65.41 MW (i.e., from bus 25 to bus 26). However, as discussed in Section A, the proposed detection method will not use the actual value of line flow to simulate the line outage. Instead, the pre-outage line flow $f_k^0$ is viewed as a variable for achieving the best match. The value of $f_k^0$ determined by the optimization problem (47) is -391.82MW (i.e., from bus 26 to bus 25). Note that the determined $f_k^0$ is not only different from the actual line flow, but the flow direction is also changed. So, according to (19), $\Delta p' = -326.84$ MW. Then, the equivalent injected powers at the terminal buses used to simulate the outage of line 25-26 are

$$\Delta p'_{25} = 326.84 \text{ MW} \quad \Delta p'_{26} = -326.84 \text{MW} \quad (53)$$

Table 2 gives the injected false data for the terminal buses of line 25-26. The second column is the injected false load data. The third column represents the total false data needed to be injected at the terminal buses, which is the sum of the injected false load and the additional power used to simulate the outage of the line. The last column is the maximum attacking amount at a load bus. It can be seen from Table 2 that the injected false load data at buses 25 and 26 are -214.84MW and 396.34MW, which are greater than the maximum attacking amount 112MW and 69.5MW, respectively. However, when the additional powers 326.84 MW, −326.84 MW are added into buses 25 and 26, the net additional power injections at buses 25 and 26 become 112MW and 69.5MW, respectively. The attacking amounts at buses 25 and 26 are no more than their maximum attacking amount, so constraints (42) and (43) are satisfied. It is further verified that all the false data injections at load buses are summed to zero and limited within $[-\tau D_d, \tau D_d]$.

TABLE 2 FALSE INJECTION DATA AT TERMINAL BUSES

| Bus | $\Delta D_d$ (MW) | $\Delta p'$ or -$\Delta p'$(MW) | $\Delta P_d$ (MW) | $\tau D_d$ (MW) |
|---|---|---|---|---|
| 25 | -214.84 | 326.84 | 112 | 112 |
| 26 | 396.34 | -326.84 | 69.5 | 69.5 |

After the bad data is injected, the residuals of all candidate lines are recalculated using (39) and shown in Table 3. Similar to Table 1, only the top 5 lowest residuals are reported. It can be seen that the residual for line 25-26 ranks the fifth. Accordingly, its outage cannot be detected based the detection method in [19-20], which requires the outaged line to have the minimum residual. In addition, constraints (9)-(11) ensure that the injected false data follow KCL and KVL and the attacking amounts are also limited within a reasonable range to avoid being detected.

TABLE 3 TOP 5 LOWEST RESIDUALS AFTER ATTACK

| rank | Line | $r_k$ |
|---|---|---|
| 1 | 6-31 | 0.0370 |
| 2 | 7-8 | 0.0372 |
| 3 | 26-29 | 0.0419 |
| 4 | 13-14 | 0.0663 |
| 5 | 25-26 | 0.0697 |

Table 4 gives the simulation results of masking the outages of 24 lines (i.e., 1-2, 1-39, 2-3, 2-25, 5-6, 5-8, 6-7, 7-8, 8-9, 9-39, 10-11, 10-13, 12-11, 14-15, 15-16, 16-17, 16-19, 16-21, 17-27, 21-22, 25-26, 26-27, 26-28 and 26-29) for different values of $\tau$. The number in the bracket ranks the corresponding residual among those of the 24 candidate lines. It can be seen that the outaged line contributes to the minimum residual value, which indicates that its outage can be detected based on the calculated residual using (17). For a greater $\tau$, the residual $r_k$ will be increased. For example, when $\tau = 0.5$, the residual of line 14-15 is 1.35; when $\tau = 1.0$, the residual is increased to 2.00. This can be explained by the optimization problem (47). As discussed, the objective of (47) is to maximize the value of the residual. As $\tau$ increases, the attacking amount at a bus becomes larger according to constraint (42)-(44). Accordingly, the feasible region of the quadratic optimization gets larger, so its objective, the residual, will become larger.

TABLE 4 RESIDUALS OF LINES FOR DIFFERENT $\tau$

| Line | $\tau = 0.5$ | $\tau = 1.0$ | $\tau = 1.5$ |
|---|---|---|---|
| 1-2 | 1.53(5) | 1.63(5) | 1.73(9) |
| 1-39 | 0.61(8) | 0.68(8) | 0.74(9) |
| 2-3 | 0.57(2) | 0.63(2) | 0.68(19) |
| 2-25 | 0.90(4) | 1.05(5) | 1.22(5) |
| 5-6 | 53.99(24) | 58.62(24) | 63.44(24) |
| 5-8 | 0.08(11) | 0.09(11) | 0.12(11) |
| 6-7 | 0.55(12) | 0.63(12) | 0.71(12) |
| 7-8 | 0.14(11) | 0.18(12) | 0.22(12) |
| 8-9 | 0.25(9) | 0.29(10) | 0.34(10) |
| 9-39 | 0.12(9) | 0.14(10) | 0.18(10) |
| 10-11 | 2.50(5) | 3.11(8) | 3.80(8) |
| 10-13 | 3.86(8) | 4.52(9) | 5.24(10) |
| 12-11 | 0.04(16) | 0.14(16) | 0.31(16) |
| 14-15 | 1.35(6) | 2.00(6) | 2.78(12) |
| 15-16 | 15.27(1) | 17.31(1) | 19.48(1) |
| 16-17 | 10.76(1) | 12.09(1) | 13.50(1) |
| 16-19 | 2.50 (12) | 2.87(13) | 3.54(13) |
| 16-21 | 0.28(1) | 0.49(1) | 0.81(1) |
| 17-27 | 0.03(14) | 0.07(13) | 0.12(10) |
| 21-22 | 1.57(1) | 2.22(1) | 2.98(1) |
| 25-26 | 0.07(5) | 0.12(8) | 0.17(20) |
| 26-27 | 0.59(1) | 0.71(1) | 0.85(1) |
| 26-28 | 0.04(13) | 0.06(14) | 0.10(14) |
| 26-29 | 0.15(19) | 0.33(19) | 0.61(19) |

We can also see from Table 4 that the injected false data can effectively mask the outages of lines. Overall, after false data are injected, the residual of line 2-3 ranks the second lowest and the lowest of the residuals of the other lines (except 15-16, 16-17, 16-21, 21-22 and 26-27) rank the 4[th]. This indicates that the outages of these lines cannot be detected any more since the line with the minimum residual is presumed to be the outaged line. Additionally, there is a general trend that the rank of the residual of the outaged line increases as $\tau$ increases. For instance, line 2-3 has the second lowest residual if $\tau = 0.5$. However, when $\tau = 1.5$ and we inject the false data $\Delta D$ determined by (47), the residual is increased to 0.74, which ranks the 19[th] lowest among the 24 candidate lines.

TABLE 5 RESIDUAL OF LINE 26-27

| Rank | $\tau = 0$ | $\tau = 0.5$ | $\tau = 1.0$ | $\tau = 1.5$ |
|---|---|---|---|---|
| 1 | 0.48 | 0.58 | 0.71 | 0.84 |
| 2 | 1.63 | 1.70 | 1.56 | 1.21 |
| 3 | 1.69 | 1.73 | 1.78 | 1.61 |

It is observed that the outages of 5 out of the 24 lines under consideration (i.e., lines 15-16, 16-17, 16-21, 21-22 and 26-27) cannot be masked. This can be also explained by the optimization problem (47). Since the attacking amount $\Delta D$ is limited within a certain range, the increase in the residual due to the false data is thus limited. So, if the minimum residual of the outaged line is much smaller than the second lowest one, then the false data might not be able to increase the residual of the outaged line such that it is greater than the second lowest one. We use line 26-27 as an example to illustrate this point. This line has the minimum residual 0.48 when $\tau = 0$, so its outage can be detected. When $\tau = 0.5$, the residual is increased to 0.58, but it is still less than the second lowest one 1.70. So, the outaged of this line can still be detected although the attacker has injected false data to disrupt the line outage detection. The same trend applies for $\tau = 1.0$ and $\tau = 1.5$. Hence, an attacker cannot inject false data to mask its outage. However, when $\tau$ is further increased to 2.0 (not shown in Table 4), the residual of line 26-27 will rank the first. Accordingly, this outage of the line can be masked by injecting the false data. It is verified that the outages of lines 16-21 and 21-22 can also be masked, when $\tau$ is increased to 3.0 and 3.5, respectively.

## V. CONCLUSIONS AND FUTURE WORK

Smart grids are subject to the high risk of cyber-attacks due to the highly integrations of communication networks. In this paper, we further propose a novel attack model to mask the outage of a single line by injecting false data into a set of measurements. A quadratic programming problem is set up to determine the optimal attack vector than can maximize the residual of the outaged line.

The success of line outage detection depends on whether the outaged line can generate the minimum residual. Consequently, if the proposed attack scheme can increase the residual significantly, such an outage cannot be detected. This motivates us to develop more secure detection approaches for line outages and develop effective defending strategies against such attacks. This is still a very challenging problem to be addressed in a future work.

In [14], we showed that an attacker only needs to obtain the network information of the attacking region instead of that of the entire power grid. This is done by making sure that the variations of phase angles of all boundary buses connected to the same island of the non-attacking region are the same. In [15], we further presented a heuristic algorithm to find an optimal attacking region which requires the reduced network information. In this paper, we assume that the attacker has the full network information of the power network. In the future work, we will investigate the possibility of launching such an attack based on local network information. We are also developing effective protection strategies against such topology attacks.

**Xuan Liu** (M'14) received the B.S. and M.S degrees from Sichuan University, China, in 2008 and 2011, respectively, both in electrical engineering. He is currently working toward the Ph.D. degree in the Electrical and Computer Engineering Department, Illinois Institute of Technology. His research interests include smart grid security, operation and economics of power systems.

**Zhiyi Li** (S' 14) received the B.S. degree from Xi'an Jiaotong University, Xi'an, China, in 2011 and the M.S. degree from Zhejiang University, China, in 2014, both in electrical engineering. He is currently pursuing the Ph.D. degree in the Electrical and Computer Engineering Department, Illinois Institute of Technology. His research interests include cyber-physical power system and power system optimization.

**Zuyi Li** (SM'09) received the B.S. degree from Shanghai Jiaotong University, Shanghai, China, in 1995, M.S. degree from Tsinghua University, China in 1998, and the Ph.D. degree from the Illinois Institute of Technology (IIT), Chicago, in 2002, all in electrical engineering. Presently, he is a Professor in the Electrical and Computer Engineering Department at IIT. His research interests include economic and secure operation of electric power systems, cyber security in smart grid, renewable energy integration, electric demand management of data centers, and power system protection.